\documentclass[a4paper,11pt]{article}
\usepackage{pos}
\usepackage{graphicx}
\usepackage{subfigure}
\usepackage{wrapfig}
\usepackage{caption}
\usepackage{mwe}
\usepackage{amsmath}
\usepackage{hyperref}

\title{Monte Carlo results on the Collins asymmetries in $e^+e^-$ annihilation}

\author*[a]{A. Kerbizi}
\author[b]{L. L\"onnblad}
\author[a]{A. Martin}

\affiliation[a]{Dipartimento di Fisica, Universitá degli Studi di Trieste and INFN Sezione di Trieste,\\
Via Valerio 2, 34127 Trieste, Italy}

\affiliation[b]{Department of Physics, Box 118, 221 00 Lund, Sweden}

\emailAdd{albi.kerbizi@ts.infn.it}

\def\GeV{\rm GeV}
\def\q{q}
\def\qp{\q'}
\def\qbar{\bar{q}}
\def\qbarp{\bar{q}'}
\def\fL{f_{\rm L}}
\def\thetaLT{\theta_{\rm LT}}

\def\phiH{\phi_{12}}
\def\n{\hat{\textbf{n}}}
\def\T{\hat{\textbf{T}}}

\def\X{z}
\def\Xa{\X_1}
\def\Xb{\X_2}
\def\Xi{\X_i}

\def\pmin{\textbf{p}_-}
\def\pp{\textbf{p}_+}

\def\kqp{k'}
\def\kVec{\textbf{k}}
\def\kbarVec{\bar{\textbf{k}}}
\def\xq{\hat{\textbf{x}}_q}
\def\yq{\hat{\textbf{y}}_q}
\def\zq{\hat{\textbf{z}}_q}
\def\xqbar{\hat{\textbf{x}}_{\qbar}}
\def\yqbar{\hat{\textbf{y}}_{\qbar}}
\def\zqbar{\hat{\textbf{z}}_{\qbar}}
\def\xu{\hat{\textbf{x}}}
\def\yu{\hat{\textbf{y}}}
\def\zu{\hat{\textbf{z}}}

\def\Iq{1^q}

\def\sigmaXq{\sigma_x^q}
\def\sigmaYq{\sigma_y^q}
\def\sigmaZq{\sigma_z^q}
\def\Iqbar{1^{\qbar}}

\def\sigmaXqbar{\sigma_x^{\qbar}}
\def\sigmaYqbar{\sigma_y^{\qbar}}
\def\sigmaZqbar{\sigma_z^{\qbar}}

\def\Trqq{\rm{Tr}_{q\qbar}}

\def\kt{\textbf{k}_{\rm T}}
\def\kpt{\textbf{k}'_{\rm T}}
\def\pt{\textbf{p}_{\rm T}}

\def\ktbar{\bar{\textbf{k}}_{\rm T}}
\def\kptbar{\bar{\textbf{k}}'_{\rm T}}
\def\Pt{\textbf{P}_{\rm T}}

\def\V{\textbf{V}}
\def\T{\textbf{T}}
\def\Tr{\rm Tr}

\def\Hq{H_{1\q}^{\perp\,h}}
\def\Hqbar{H_{1\qbar}^{\perp\,h}}
\def\AOneTwo{A_{12}}
\def\AOneTwoUL{A_{12}^{\rm UL}}
\def\AOneTwoUC{A_{12}^{\rm UC}}
\def\Azero{A_0}
\def\AzeroUL{A_{0}^{\rm UL}}
\def\AzeroUC{A_{0}^{\rm UC}}

\def\P{\textbf{P}}
\def\PTzeroVec{\P_{\rm 0\,T}}
\def\PTzero{P_{\rm 0\,T}}

\abstract{The quark spin effects have been introduced in the Pythia 8 Monte Carlo event generator for the simulation of $e^+e^-$ annihilation by interfacing the generator with the StringSpinner package. The package allows to simulate quark spin effects in the string fragmentation routine of Pythia by using the string+${}^3P_0$ model, recently applied to the fragmentation of a string stretched between a quark-antiquark pair with correlated spin states.

StringSpinner is used to carry out simulations of $e^+e^-$ annihilation at the center of mass energy of $10.6\,\GeV$. The Collins asymmetries are extracted from the simulated data for back-to-back pion pairs using both the thrust axis method and the hadronic plane method. The results are compared with the data from the BELLE and BABAR experiments, finding a satisfactory agreement.}

\FullConference{7th International Workshop on “Transverse phenomena in hard processes and the transverse structure of the proton (Transversity2024)\\
 3-7 June 2024\\
Trieste, Italy\\}

\begin{document}
\maketitle

\section{Introduction}
The annihilation reaction $e^+e^-\rightarrow \rm hadrons$ is an important process to study hadronization, namely the conversion of quarks and gluons in hadrons. According to the factorization theorem~\cite{Collins:1981uk}, the reaction can be factorized in an elementary hard interaction, $e^+e^-\rightarrow \q\qbar$, where a quark pair is produced, and in the hadronization of $\q$ and $\qbar$ in the final state hadrons. The latter is usually described by fragmentation functions (FFs). A particularly important FF is the Collins function $\Hq$, which describes the fragmentation of a transversely polarized quark $q$ in an unpolarized hadron $h$~\cite{Collins:1992kk}.
In the annihilation reaction $e^+e^-\rightarrow h_1\,h_2\,X$, the Collins effect is responsible for the correlations between the azimuthal angles of the hadrons $h_1$ and $h_2$ produced back-to-back in the $e^+e^-$ center of mass (CM) system. The strength of the correlations is quantified by the Collins asymmetry in $e^+e^-$, which couples the Collins function $\Hq$ for the fragmentation of $\q$ with the function $\Hqbar$ for the fragmentation of $\qbar$. The Collins asymmetry in $e^+e^-$ has been measured to be non-vanishing by the BELLE~\cite{Belle:2008fdv,Belle:2019nve}, BABAR~\cite{BaBar:2013jdt,BaBar:2015mcn} and BESIII 
\cite{BESIII:2015fyw} experiments.


The Collins FF allows also to access the partonic transverse spin structure of the nucleons. An example is the extraction of the transverse spin distribution of quarks in a transversely polarized nucleon described by the transversity parton distribution function (PDF) $h_1^q$, which requires the knowledge of the Collins FF. The transversity PDF and the Collins FF have been extracted by different groups by a combined phenomenological analysis of the Collins asymmetries in SIDIS and in the $e^+e^-$ annihilation (for a review see, e.g., Ref.~\cite{Anselmino:2020vlp}).

An alternative approach to the phenomenological extractions of the Collins FF is the modeling of hadronization and its implementation in Monte Carlo event generators. This was done in Refs. ~\cite{Kerbizi:2021gos,Kerbizi:2021pzn,Kerbizi:2023cde} for the SIDIS process. Now we have implemented the quark spin effects in the Pythia 8.3 event generator~\cite{Bierlich:2022pfr} for the simulation of $e^+e^-$ annihilation to hadrons. The spin effects in hadronization are introduced by using the string+${}^3P_0$ model described in Ref.~\cite{Kerbizi:2023luv}, which extended the model of Ref.~\cite{Kerbizi:2021gos} to the string fragmentation process of a $\q\qbar$ pair with entangled spin states. The implementation of the model in Pythia is achieved by further developing the StringSpinner package~\cite{Kerbizi:2021pzn,Kerbizi:2023cde} to handle $e^+e^-$ events. A complete description of the work can be found in Ref. \cite{Kerbizi:2024vpd}. In Sec. \ref{sec:implementation} of these proceedings we recall the implementation of the string+${}^3P_0$ model in Pythia for $e^+e^-$ annihilation. The new StringSpinner package is used to simulate $e^+e^-$ events at CM energy $\sqrt{s}=10.6\,\GeV$, corresponding to the energy of the BELLE and BABAR experiments, and the resulting Collins asymmetries for back-to-back hadrons from the simulated events are compared with the data in Sec. \ref{sec:results}. The conclusions are given in Sec. \ref{sec:conclusions}.

\section{Implementation of the spin effects in Pythia for $e^+e^-$}\label{sec:implementation}
The $e^+e^-$ process is considered at leading order as mediated by the exchange of a virtual photon. Gluon radiation described in Pythia by the final state parton shower has been switched off.

\begin{figure}[tb]
\centering
\begin{minipage}[b]{0.45\textwidth}
\hspace{-2.0em}
\includegraphics[width=1.0\textwidth]{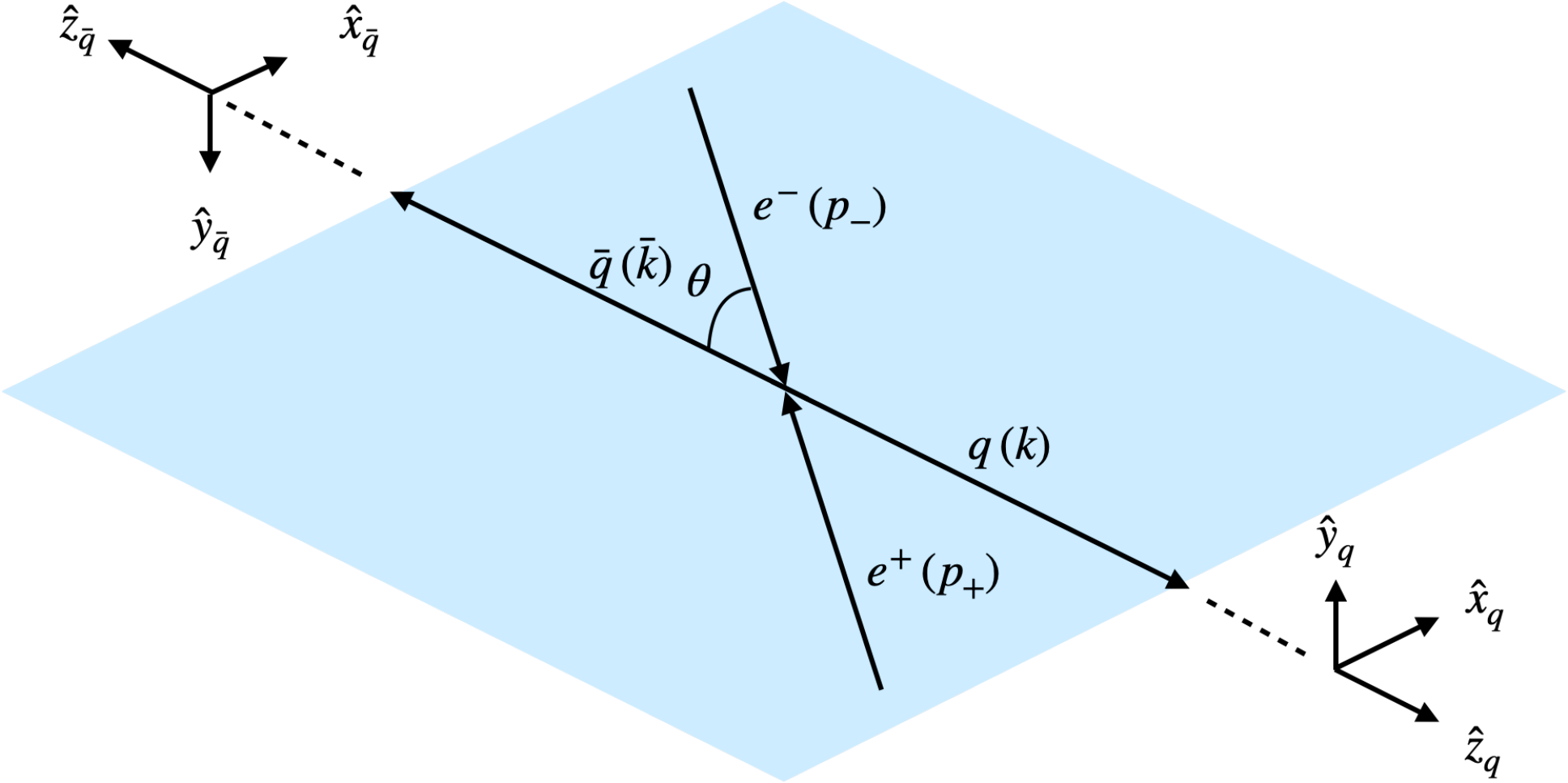}
\end{minipage}
\begin{minipage}[b]{0.45\textwidth}
\hspace{1.0em}
\includegraphics[width=1.0\textwidth]{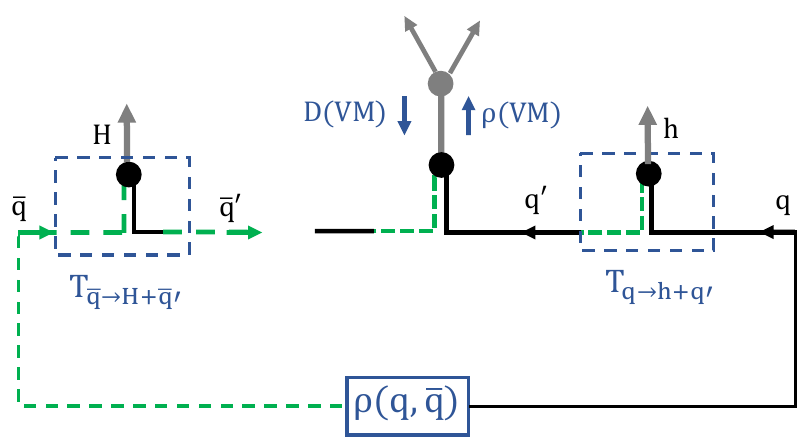}
\end{minipage}
\caption{Kinematics of the $e^+e^-\rightarrow q\qbar$ in the CM system (left). Representation of the polarized string fragmentation process in StringSpinner (right).}
\label{fig:kinematics}
\end{figure}

To begin the simulation, we let Pythia generate the kinematics of the hard reaction $e^+e^-\rightarrow \q\qbar$, shown in Fig. \ref{fig:kinematics}a. The angle $\theta$ is angle between the momentum $\pmin$ of $e^-$ and the momentum $\kVec$ of $\q$. The momenta of $e^+$ and $\qbar$ are indicated by $\pp$ and $\kbarVec$, respectively. Following Ref.~\cite{Kerbizi:2023luv}, we introduce the quark helicity frame (QHF) by the axes $\lbrace \xq, \yq,\zq\rbrace$ defined by $\zq=\kVec/|\kVec|$, $\yq=\pmin\times\zq/|\pmin\times \zq |$ and $\xq= \yq\times \zq$. The antiquark helicity frame (AHF), defined by the axes $\lbrace \xqbar, \yqbar, \zqbar\rbrace$, is obtained analogously to the QHF by replacing $\kVec$ with $\kbarVec$. The QHF and AHF are also shown in Fig. \ref{fig:kinematics} (left).

Before starting the fragmentation of the string stretched between $\q$ and $\qbar$, the joint spin density matrix~\cite{Kerbizi:2023luv}
\begin{equation}\label{eq:rho}
\rho(\q,\qbar)=\frac{1}{4}\times \left[\Iq\otimes \Iqbar - \sigmaZq\otimes \sigmaZqbar + \frac{\sin^2\theta}{1+\cos^2\theta}\,(\sigmaXq\otimes \sigmaXqbar + \sigmaYq\otimes \sigmaYqbar)\right],
\end{equation}
is set up, where the quark mass has been neglected. The matrix $\sigma_{i}^{\q(\qbar)}$ indicates the Pauli matrix along the axis $i=\xu,\yu,\zu$ in the QHF (AHF). The joint spin density matrix implements the correlations between the spin states of $\q$ and $\qbar$, and in particular the correlations between the transverse spins of $\q$ and $\qbar$. 

Pythia then starts the fragmentation of the $\q-\qbar$ string by selecting randomly emissions of hadrons from the $\q$ and $\qbar$ sides of the string. As can be seen in Fig. \ref{fig:kinematics} (right), emissions from the $\q$ side are viewed as splittings $\q\rightarrow h + \qp$, where $h$ is the emitted hadron with four momentum $p$ and $\qp$ is the leftover quark with four-momentum $\kqp$. The transverse momenta of $\q$, $h$ and $\qp$ with respect to the string axis are defined to be $\kt$, $\pt$ and $\kpt$, respectively. Momentum conservation implies $\kpt=\kt-\pt$.
Analogously, emissions from the $\qbar$ side are viewed as splittings $\qbar\rightarrow H + \qbarp$, with $H$ being the emitted hadron and $\qbarp$ the leftover antiquark. The transverse momenta of $\qbar$, $H$ and $\qbarp$ with respect to the string axis are defined as $\ktbar$, $\Pt$ and $\kptbar$, respectively. They are related by $\kptbar = \ktbar-\Pt$. 

Following the previous implementation of StringSpinner~\cite{Kerbizi:2023cde} only the production of pseudoscalar (PS) mesons and vector mesons (VMs) is allowed. Then, if the splitting is taken from the $\q$ side, the hadron $h$ is accepted with a probability
\begin{equation}\label{eq:wh}
w_h(\pt;\kt)=\Trqq\left[\T_{\qp,h\,\q}\,\rho(\q,\qbar)\T^{\dagger}_{\qp,h\,\q}\right]/\Trqq\left[\T_{\qp,h\,\q}\,\T^{\dagger}_{\qp,h\,\q}\right],
\end{equation}
where $\T_{\qp,h,\q}=T_{\qp,h,\q}\otimes \Iqbar$ and $T_{\qp,h,\q}$ is the splitting matrix of the string+${}^3P_0$ model~\cite{Kerbizi:2023luv} describing the splitting $\q\rightarrow h+\qp$. Equation (\ref{eq:wh}) modifies the azimuthal distribution of $h$ produced by Pythia in agreement with the rules of the string+${}^3P_0$ model. 

If $h$ is a VM, its decay is performed as in Ref.~\cite{Kerbizi:2023luv}, using the spin density matrix $\rho_{aa'}(h)\propto \Trqq\left[\T^{a}_{\qp,h\,\q}\,\rho(\q,\qbar)\T^{a'\,\dagger}_{\qp,h\,\q}\right]$, where the splitting amplitude for VM emission is written as $\T_{\qp,h,\q}=\T^a_{\qp,h,\q}\,\V^*_a$ and $\V_a$ is the linear polarization vector of the VM expressed in the QHF. Once $h$ is accepted, the spin correlations are propagated by updating the joint spin density matrix $\rho(\q,\qbar)\rightarrow \rho(\qp,\qbar)\propto \T^{a}_{\qp,h,\q}\,\rho(\q,\qbar)\T^{a'\dagger}\,D_{a'a}$ of the $\qp\qbar$ pair. For a VM emission $D_{a'a}$ is the decay matrix implementing the decay process of the VM~\cite{Kerbizi:2021gos}, as required by the Collins-Knowles algorithm~\cite{Collins:1987cp,Knowles:1988vs} (Fig. \ref{fig:kinematics}, right). For a PS meson emission the indices $a$ and $a'$, and $D_{a'a}$ are removed.

For a splitting from the $\qbar$ side the procedure is analogous, but the splitting amplitude $\T_{\qbarp,H,\qbar}=\Iq\otimes T_{\qbar',H,\qbar}$ is used, with $T_{\qbarp,H,\qbar}$ being the antiquark splitting matrix~\cite{Kerbizi:2023luv}. If $H$ is emitted, e.g., after the emission of $h$ from the $\q$ side, $H$ is accepted with the probability
\begin{equation}\label{eq:wH}
w_H(\Pt;\ktbar)=\Tr_{\qp\qbar}\left[\T_{\qbarp,H\,\qbar}\,\rho(\qp,\qbar)\T^{\dagger}_{\qbarp,H\,\qbar}\right]/\Tr_{\rm \qp\qbar}\left[\T_{\qbar',H\,\qbar}\,\T^{\dagger}_{\qbarp,H\,\qbar}\right].
\end{equation}
The probability for emitting $H$ is now conditional to the emission of $h$ from $\q$, due to the fact that in Eq. (\ref{eq:wH}) enters the joint spin density matrix $\rho(\qp,\qbar)$. This results in a correlation between the azimuthal angles of the transverse momenta $\pt$ and $\Pt$.
For the decay of VMs as well as the propagation of the spin correlations, the same procedure as for the $\q$ splitting above is followed provided that the replacement $\T_{\qp,h,\q}\rightarrow \T_{\qbarp,H,\qbar}$ is performed. The four-momenta of hadrons emitted from the $\qbar$ side are expressed in the AHF.

The described procedure is applied until the exit condition of the string fragmentation process is called by Pythia and the process is terminated.

The free parameters of the string+${}^3P_0$ model responsible for the spin effects are the complex mass $\mu=\rm{Re}(\mu)+i\,\rm{Im}(\mu)$ implementing the ${}^3P_0$ mechanism of quark pair production at string breaking, the fraction of longitudinally polarized VMs $\fL$ and $\thetaLT$ allowing for the oblique polarization of VMs~\cite{Kerbizi:2021gos}.

\section{Simulation results on Collins asymmetries}\label{sec:results}
\subsection{The simulated data sample}
Using the new StringSpinner package, we simulated $60\times 10^6$ $e^+e^-$ events at the CM energy $\sqrt{s}=10.6\,\GeV$ of the BELLE and BABAR experiments. 
The annihilation reaction is mediated by a virtual photon, which is allowed to decay to $\q\qbar$ pairs with $q=u,d,s$. The production of heavier quarks has been switched off. The parameter settings used for the simulation are the same as those in Ref.~\cite{Kerbizi:2023cde}, except for $\fL$ and $\thetaLT$. In this work, we have used $\fL=0.12$ and $\thetaLT=-0.65$. These values have been chosen to reproduce the $e^+e^-$ experimental results, and are found to give satisfactory results also for the SIDIS observables.

For each simulated $e^+e^-$ event the thrust axis is evaluated by using the Pythia routine. As in the experimental data analyses~\cite{Belle:2008fdv,BaBar:2013jdt,BaBar:2015mcn}, the thrust axis is defined as the normalized vector $\n$ that maximises the event variable $T=\sum_j\,|\P_j\cdot\n|/\sum_j\,|\P_j|$, where $\P_j$ is the momentum of the hadron $h_j$ in the CM system and the index $j$ runs over the final state hadrons. $T$ is referred to as the thrust and, as in the data analysis of the experiments, we ask $T>0.8$ to select two-jet like events.
Then pairs of hadrons $h_1$ and $h_2$ produced in the same event are formed and stored in the simulated data sample. The hadrons are requested to be produced in two opposite hemispheres (i.e. almost back-to-back), by requiring $(\P_1\cdot \n)\,(\P_2\cdot \n)<0$. To reduce the association of hadrons to the wrong hemisphere, the photon transverse momentum $Q_{\rm T}$ evaluated in the rest frame of the $h_1\,h_2$ pair is required to be $Q_{\rm T}<3.5\,\GeV/c$.

The simulated data sample has been analysed following the procedure used in the BELLE and BABAR experiments~\cite{Belle:2008fdv,BaBar:2013jdt,BaBar:2015mcn} and the Collins asymmetries have been extracted both with the thrust axis method ($\AOneTwo$) and hadronic plane method ($\Azero$) as described in the following.

\subsection{The $\AOneTwo$ asymmetries}
To extract the $\AOneTwo$ Collins asymmetries we use the distribution of the azimuthal angle $\phiH = \phi_1+\phi_2$, with $\phi_i$ being the azimuthal angle of the hadron $i=1,2$ defined with respect to the plane formed by the beam $e^-$ and the axis $\n$. 
In this work we compare the simulation results with data from Refs. \cite{Belle:2008fdv,BaBar:2013jdt,BaBar:2015mcn} and therefore we use the true $\q\qbar$ axis as axis $\n$. The results using the thrust axis can be found in Ref. \cite{Kerbizi:2024vpd}.
The distribution of the produced back-to-back hadrons $h_{1}h_{2}$ is expected to be 
\begin{equation}\label{eq:dN}
    N_{12}(\phiH;z_1\,z_2)\propto 1+\frac{\langle \sin^2\theta\rangle}{\langle 1+\cos^2\theta\rangle}\,A_{12}(z_1,z_2)\,\cos\phiH,
\end{equation}
where we have introduced the fractional energy $z_i=2\,E_i/\sqrt{s}$ of the hadron $h_i$, with $i=1,2$ and $E_i$ being the energy of $h_i$. In each kinematic bin the angular distribution in Eq. (\ref{eq:dN}) is used to construct the normalized yield $R_{12}(\phiH;\Xa,\Xb)=N_{12}(\phiH;\Xa,\Xb)/\langle N_{12}\rangle$, where $\langle N_{12}\rangle$ is the average yield in the considered two-dimensional $\Xa\times \Xb$ bin. Finally, normalized yields are calculated for unlike (U), like (L) and charged (C) pairs. The ratio $R_{12}^{UL(UC)}=R_{12}^U/R_{12}^{L(C)}$ has a similar expression as in Eq. (\ref{eq:dN}), with the amplitude of the $\cos\phiH$ modulation being given by $A_{12}^{UL(UC)}\simeq A_{12}^{U}-A_{12}^{L(C)}$, namely the difference between the Collins asymmetry for unlike charge and like charge (charged) hadrons. 

\begin{figure}[tbh]
\centering
\begin{minipage}[b]{0.49\textwidth}
\includegraphics[width=1.0\textwidth]{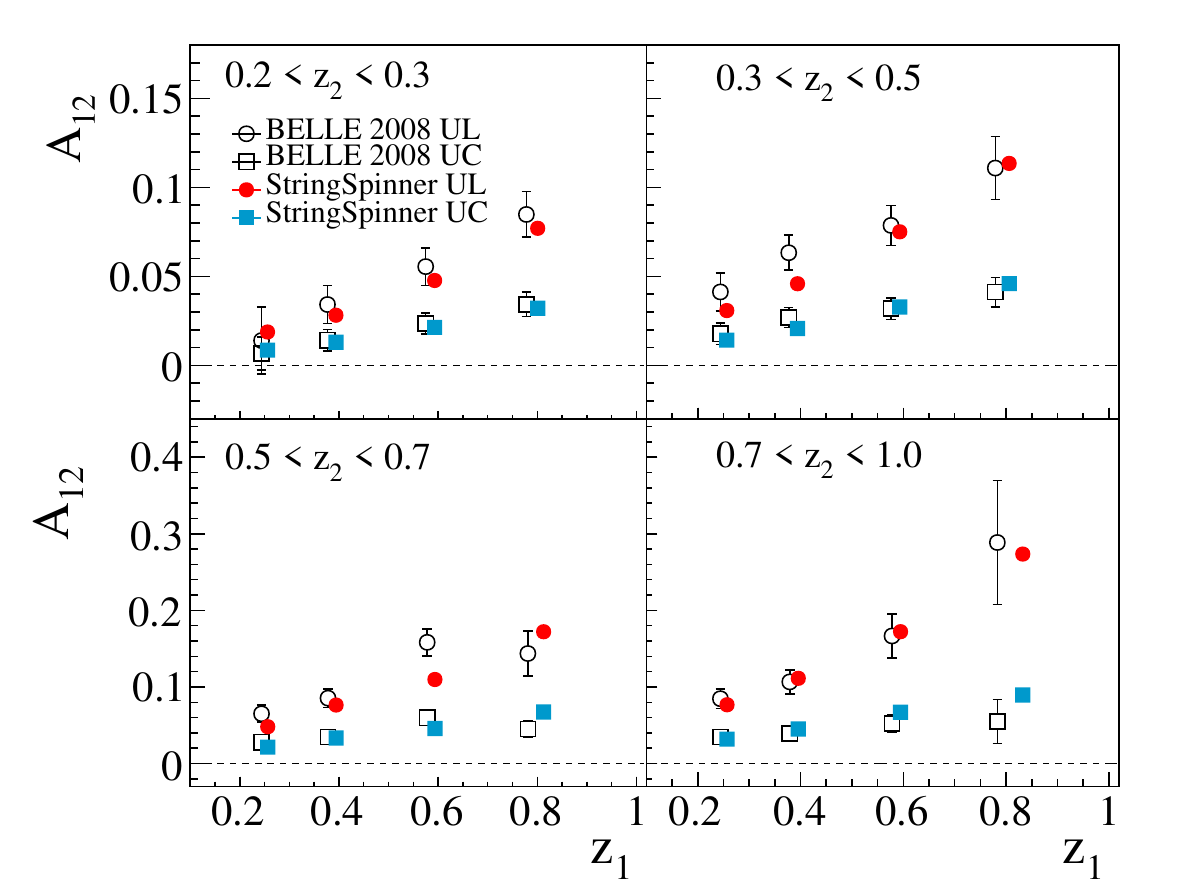}
\end{minipage}
\begin{minipage}[b]{0.49\textwidth}
\hspace{1.0em}
\includegraphics[width=1.0\textwidth]{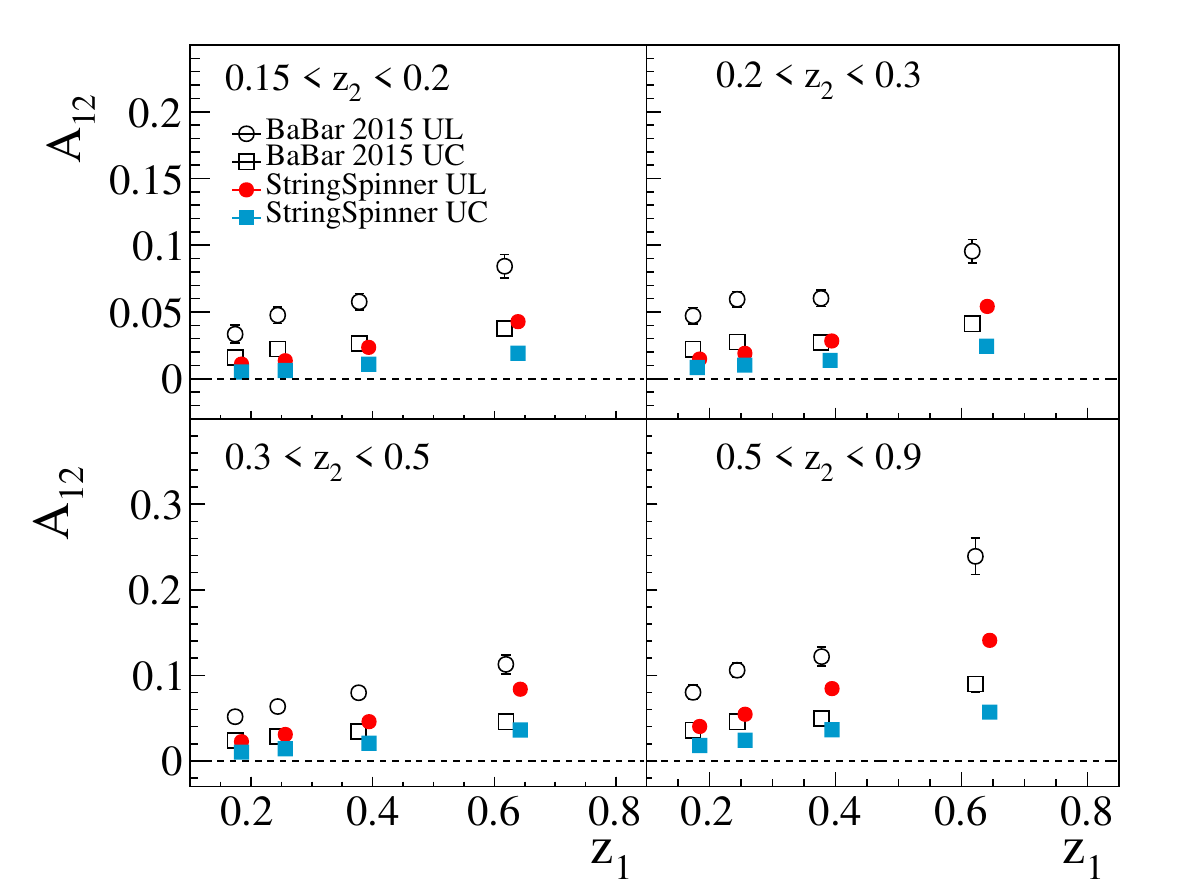}
\end{minipage}
\caption{Comparison of the asymmetries $\AOneTwoUL$ (circles) and $\AOneTwoUC$ (rectangles) for pions simulated with StringSpinner (full points) with the experimental results (empty points) from BELLE~\cite{Belle:2008fdv} (left plot) and BABAR \cite{BaBar:2015mcn} (right plot).}
\label{fig:A12}
\end{figure}

Figure \ref{fig:A12} shows the StringSpinner results (full points) for the $\AOneTwoUL$ and $\AOneTwoUC$ asymmetries calculated for charged pions as a function of $z_1$ for different bins of $z_2$. In the left plot the binning is the same as in the BELLE analysis \cite{Belle:2008fdv} while in the right plot it is the same as in the BABAR analysis \cite{BaBar:2015mcn}. In the BABAR case, the additional cut on the opening angle of each hadron with respect to $\n$ being less than $\pi/4$ is applied. 
As can be seen, the asymmetries increase as a function of the fractional energy reaching values up to $30\%$ for $z_i\sim 0.8$. The $\AOneTwoUL$ asymmetries are larger as compared to the $\AOneTwoUC$ asymmetries, as they involve different mixtures of favoured and unfavoured fragmentations.

In the same figure, the experimental results (open points) by BELLE \cite{Belle:2008fdv} (left) and BABAR \cite{BaBar:2015mcn} (right) are shown. It can be seen that StringSpinner reproduces the trend and the size of both $\AOneTwoUL$ and $\AOneTwoUC$ asymmetries measured by BELLE. The simulated results are instead somewhat lower as compared to the BABAR results, which are larger than the BELLE results in the common $z_1\times z_2$ bins.
This is not the case for the simulated asymmetries, which are the same in the common bins for BELLE and BABAR, as expected from the similar steps applied in the two analyses.

Using the simulated data, the $\AOneTwo$ asymmetry has been evaluated also for back-to-back charged kaon pairs and a satisfactory agreement with the BABAR data is found. The result can be found in Ref. \cite{Kerbizi:2024vpd}.

\subsection{The $\Azero$ asymmetries}
To evaluate the asymmetry $\Azero$, the plane containing the momentum $\pmin$ of $e^-$ and the momentum $\P_2$ of $h_2$ is considered.
The plane is used to measure the azimuthal angle $\phi_0$ of the transverse momentum $\PTzeroVec$ of $h_1$ with respect to $\P_2$.
The distribution of the azimuthal angle $\phi_0$ of $h_1$ is expected to be 
\begin{eqnarray}\label{eq:N0}
     N_{0}(\phi_0;z_1,z_2,\PTzero)\propto 1+\frac{\langle\sin^2\theta_2\rangle}{\langle 1+\cos^2\theta_2\rangle}\,A_{0}(z_1,z_2,\PTzero)\,\cos 2\phi_0,
\end{eqnarray}
where $\theta_2$ is the angle between $\P_2$ and the beam $\pmin$. The amplitude $A_0$ depends on $z_1$, $z_2$, the magnitude $\PTzero$ of $\PTzeroVec$, and $Q_{\rm T}$~\cite{Boer:1997mf}. The latter dependence is not considered here.
As done for the $\AOneTwo$ asymmetry, in a given kinematic bin the angular distribution in Eq. (\ref{eq:N0}) is used to construct the normalized yield $R_{0}(\phi_0;z_1,z_2,\PTzero)=N_{0}(\phi_0;z_1,z_2,\PTzero)/\langle N_{0} \rangle$, where $\langle N_{0}\rangle$ is the average yield in the considered kinematic bin. Then, normalized yields are calculated for U, L and C pairs. The ratio $R_{0}^{UL(UC)}=R_{0}^U/R_{0}^{L(C)}$ has the same $\phi_0$-dependence as that in Eq. (\ref{eq:N0}), with the amplitude of the $\cos\phi_0$ modulation given by $A_{0}^{UL(UC)}\simeq A_{0}^{U}-A_{0}^{L(C)}$.

\begin{figure}[tbh]
\centering
\begin{minipage}[b]{0.49\textwidth}
\includegraphics[width=1.0\textwidth]{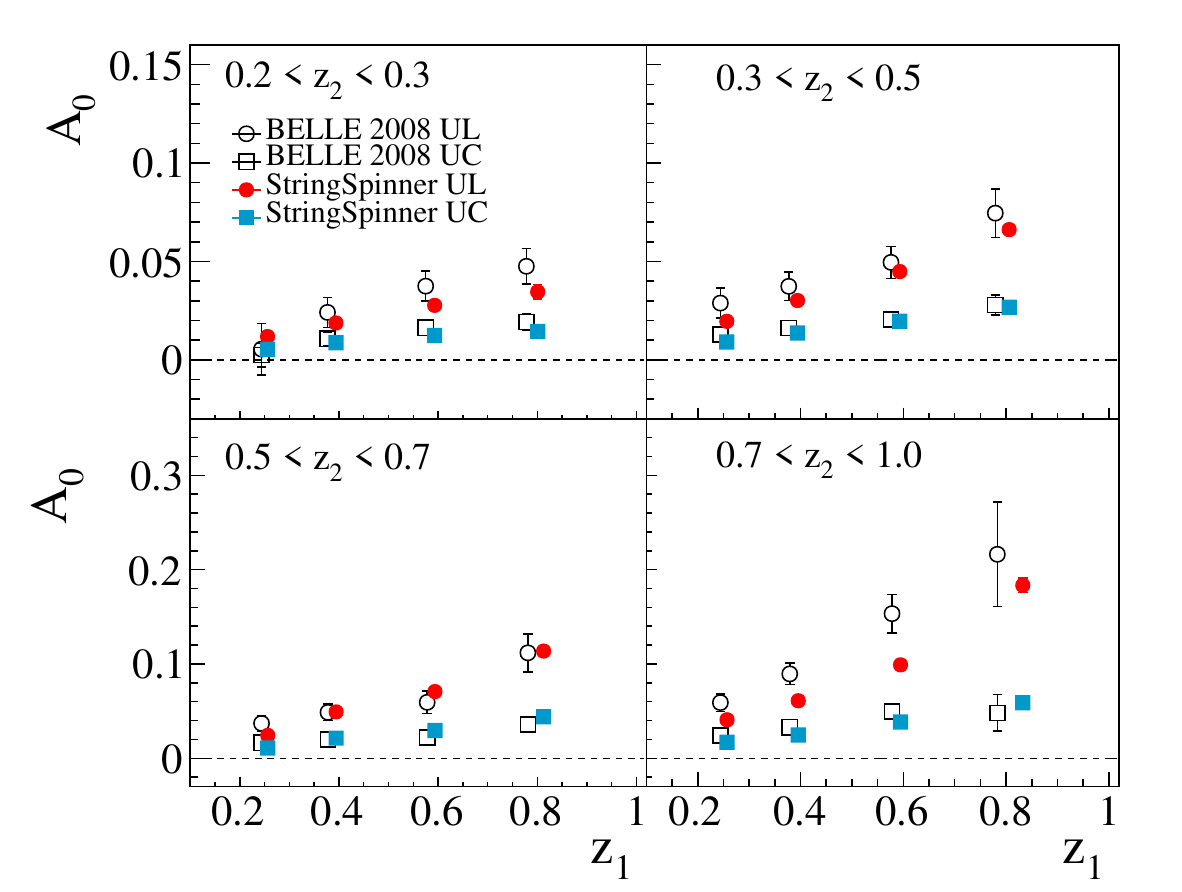}
\end{minipage}
\begin{minipage}[b]{0.49\textwidth}
\hspace{1.0em}
\includegraphics[width=1.0\textwidth]{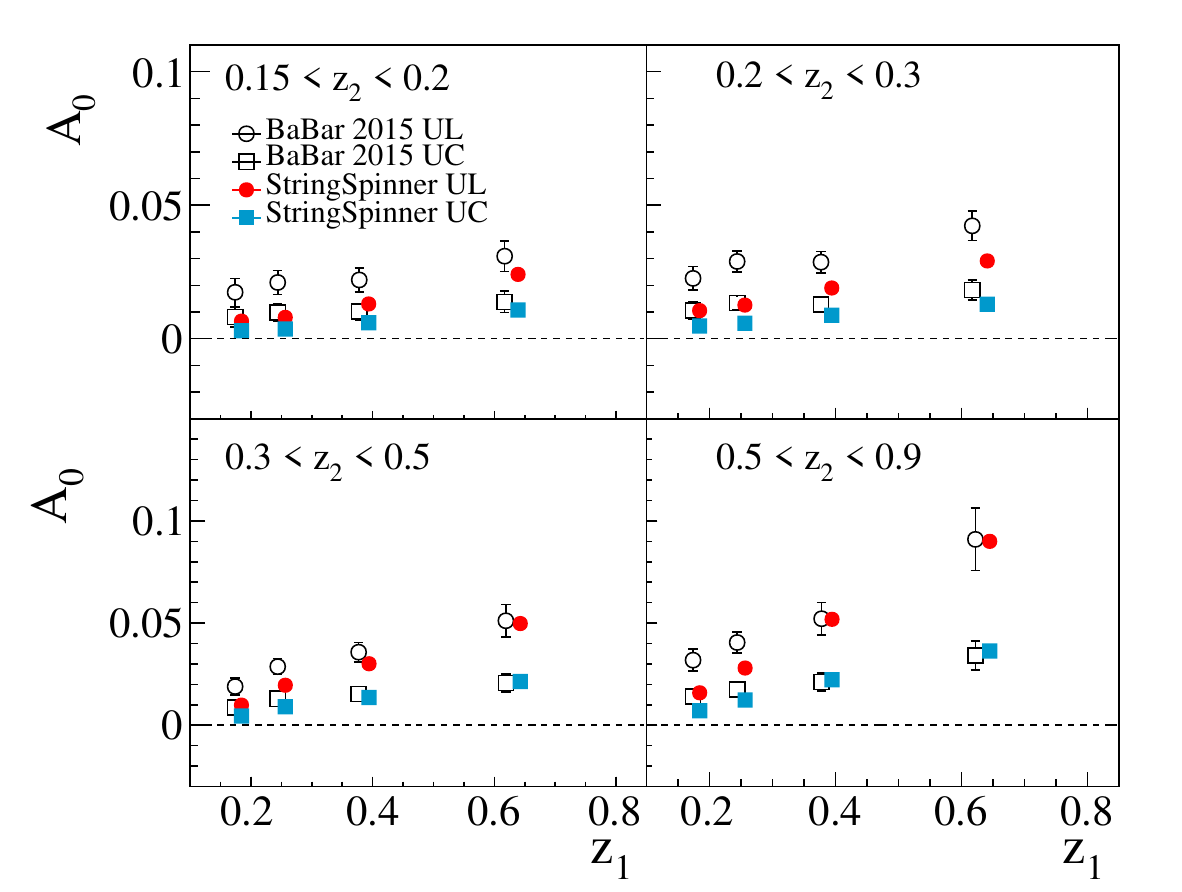}
\end{minipage}
\caption{Comparison of the asymmetries $\AzeroUL$ (circles) and $\AzeroUC$ (rectangles) for pions simulated with StringSpinner (full points) with the experimental results (empty points) from BELLE~\cite{Belle:2008fdv} (left plot) and BABAR \cite{BaBar:2015mcn} (right plot).}
\label{fig:A0}
\end{figure}

The StringSpinner results on the $\Azero$ asymmetry for back-to-back charged pions are shown in Fig. \ref{fig:A0} with different $z_1\times z_2$ binning to compare with experimental results. The asymmetry has a rising trend with $z$, and it is larger for the UL pairs than for UC pairs. Comparing with the left plot in Fig. \ref{fig:A12}, it can be seen that the simulated $\Azero$ asymmetries are smaller than the $\AOneTwo$ asymmetries. This is due to the smearing induced by the use of $\P_2$ in the $\Azero$ case instead of the true quark-antiquark axis. 

The corresponding asymmetries as measured by BELLE \cite{Belle:2008fdv} and by BABAR \cite{BaBar:2015mcn} are shown in Fig. \ref{fig:A0} by the open points. As can be seen, the simulations reproduce the measured $\Azero$ asymmetries for both experiments. Exceptions are the $\AzeroUL$ asymmetry in the highest $z$ bin when comparing with BELLE and for $z<0.3$ when comparing with BABAR.

The $\Azero$ asymmetry has been evaluated with the simulated data also for back-to-back charged kaon pairs, finding a satisfactory comparison with the BABAR data. The result can be found in Ref. \cite{Kerbizi:2024vpd}.

\begin{figure}[h]
\centering
\begin{minipage}[b]{0.5\textwidth}
\includegraphics[width=1.0\textwidth]{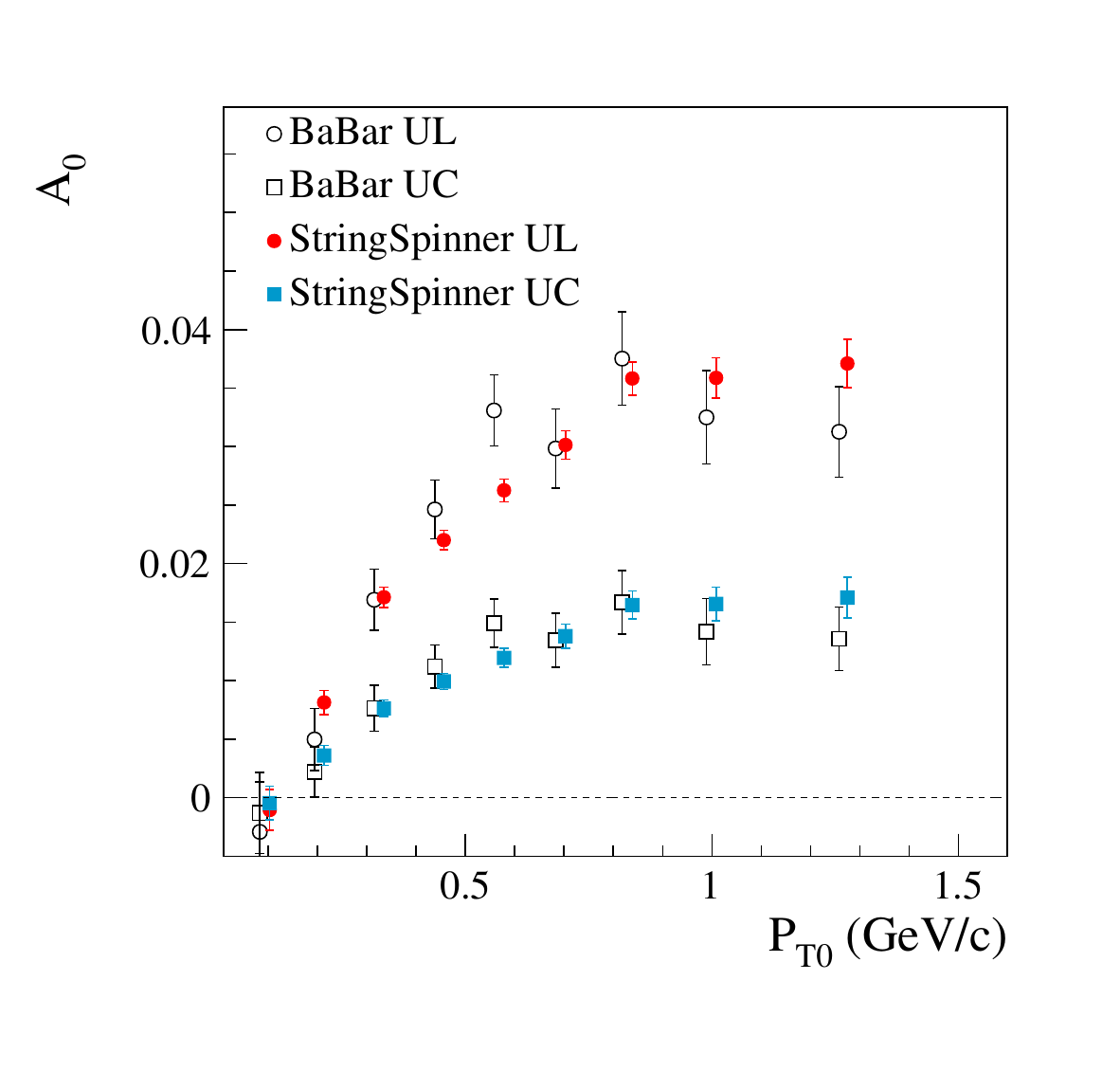}
\end{minipage}
\vspace{-2em}
\caption{Comparison between the Collins asymmetries $\AzeroUL$ (full circles) and $\AzeroUC$ (full squares) as a function of $\PTzero$ for back-to-back charged pion pairs obtained with StringSpinner (full points) and the asymmetries measured by BABAR \cite{BaBar:2013jdt} (open points).}
\label{fig:A0 PT0 BaBar 2014}
\end{figure}

Finally in Fig. \ref{fig:A0 PT0 BaBar 2014} the simulated $\Azero$ asymmetry (full points) for charged pion pairs as a function of the transverse momentum $\PTzero$ is shown. 
Both $\AzeroUL$ and $\AzeroUC$ asymmetries show a rising trend as a function of $\PTzero$ up to $\PTzero\approx 0.7\,\GeV/c$, and saturate for larger values of the transverse momentum. In the figure the corresponding $\Azero$ asymmetries measured by BABAR \cite{BaBar:2013jdt} (open points) are also shown. As can be seen, both the size and the trend of the measured asymmetries are remarkably well reproduced by the simulations. This confirms that the string+${}^3P_0$ model satisfactorily reproduces transverse spin and transverse momentum effects seen in $e^+e^-$ annihilation.

\section{Conclusions}\label{sec:conclusions}
The string+${}^3P_0$ model of hadronization has been implemented in the Pythia 8 event generator for the simulation of the $e^+e^-$ annihilation to hadrons with quark spin effects. This is achieved by using the recent application of the string+${}^3P_0$ model to the fragmentation of a string stretched between a quark-antiquark pair with entangled spin states and extending the StringSpinner package. Using the new package, $e^+e^-$ annihilation events have been simulated at the CM energy $\sqrt{s}=10.6\,\GeV$, assuming that the annihilation occurs by the exchange of a virtual photon.

The Collins asymmetries for back-to-back hadrons have been calculated from the simulated data using both the thrust axis method and the hadronic plane method. The results are compared to the data by the BELLE and BABAR experiments, finding an overall satisfactory agreement.

The obtained results confirm the soundness of the string+${}^3P_0$ model of hadronization and motivate further developments of the model.

\subsection*{Acknowledgments}
The work of AK was supported by the Ministry of University and Research (MUR) within the POLFRAG project, CUP n. J97G22000510001.

\end{document}